\documentclass[aip, apl,reprint]{revtex4-1}

\usepackage{graphicx}
\usepackage{dcolumn}
\usepackage{bm}

\begin{document}

\title[]{A widely tunable 10-$\mu$m quantum cascade laser phase-locked to a state-of-the-art mid-infrared reference for precision molecular spectroscopy}

\author{P.L.T. Sow}
\affiliation{CNRS, UMR 7538, LPL, 93430 Villetaneuse, France}
\affiliation{Universit\'{e} Paris 13, Sorbonne Paris Cit\'{e}, Laboratoire de Physique des Lasers, 93430 Villetaneuse, France}

\author{S. Mejri}
\author{S.K. Tokunaga}
\affiliation{Universit\'{e} Paris 13, Sorbonne Paris Cit\'{e}, Laboratoire de Physique des Lasers, 93430 Villetaneuse, France}
\affiliation{CNRS, UMR 7538, LPL, 93430 Villetaneuse, France}

\author{O. Lopez}
\affiliation{CNRS, UMR 7538, LPL, 93430 Villetaneuse, France}
\affiliation{Universit\'{e} Paris 13, Sorbonne Paris Cit\'{e}, Laboratoire de Physique des Lasers, 93430 Villetaneuse, France}

\author{A. Goncharov}
\altaffiliation[Permanent address: ]{Institute of Laser Physics of SB RAS, Pr. Lavrentyeva 13/3, Novosibirsk, 630090 Russia}
\affiliation{Universit\'{e} Paris 13, Sorbonne Paris Cit\'{e}, Laboratoire de Physique des Lasers, 93430 Villetaneuse, France}
\affiliation{CNRS, UMR 7538, LPL, 93430 Villetaneuse, France}

\author{B. Argence}
\author{C. Chardonnet}
\affiliation{CNRS, UMR 7538, LPL, 93430 Villetaneuse, France}
\affiliation{Universit\'{e} Paris 13, Sorbonne Paris Cit\'{e}, Laboratoire de Physique des Lasers, 93430 Villetaneuse, France}

\author{A. Amy-Klein}
\author{C. Daussy}
\affiliation{Universit\'{e} Paris 13, Sorbonne Paris Cit\'{e}, Laboratoire de Physique des Lasers, 93430 Villetaneuse, France}
\affiliation{CNRS, UMR 7538, LPL, 93430 Villetaneuse, France}

\author{B. Darqui\'{e}}
\email{benoit.darquie@univ-paris13.fr}
\affiliation{CNRS, UMR 7538, LPL, 93430 Villetaneuse, France}
\affiliation{Universit\'{e} Paris 13, Sorbonne Paris Cit\'{e}, Laboratoire de Physique des Lasers, 93430 Villetaneuse, France}

\begin{abstract}
We report the coherent phase-locking of a quantum cascade laser (QCL) at 10-$\mu$m to the secondary frequency standard of this spectral region, a CO$_2$ laser stabilized on a saturated absorption line of OsO$_4$. The stability and accuracy of the standard are transferred to the QCL resulting in a line width of the order of 10~Hz, and leading to our knowledge to the narrowest QCL to date. The locked QCL is then used to perform absorption spectroscopy spanning 6~GHz of NH$_3$ and methyltrioxorhenium, two species of interest for applications in precision measurements.
\end{abstract}

\maketitle

With their rich internal structure, molecules can play a decisive role in precision tests of fundamental physics. They are being used to test fundamental symmetries such as parity~\cite{Daussy1999, DeMille2008, Darquie2010} or parity and time reversal~\cite{Baron2014}, to measure absolute values of fundamental constants~\cite{Koelemeij2007, Lemarchand2013, Moretti2013} and their possible temporal variation~\cite{Hudson2006, Shelkovnikov2008, Truppe2013}. Many of these experiments can be cast as measurements of resonance frequencies of molecular transitions, for which ultra-stable and accurate sources in the mid-infrared (mid-IR) are highly desirable, since most rovibrational transitions are to be found in that region.

Our group has a long-standing interest in performing spectroscopic precision measurements on molecules at extreme resolutions around $10~\mu$m~\cite{Daussy1999, Ziskind2002, Shelkovnikov2008}. We are currently working on two such measurements: the determination of the Boltzmann constant, $k_{\mathrm{B}}$, by Doppler spectroscopy of ammonia~\cite{Lemarchand2011, Lemarchand2013} and the first observation of parity violation by Ramsey interferometry of a beam of chiral molecules~\cite{Darquie2010, Tokunaga2013}. For these experiments, we currently use spectrometers based on custom built ultra-stable CO$_2$ lasers. We obtain the required metrological frequency stability and accuracy --- 10~Hz line width, 1~Hz stability at 1~s, accuracy of a few tens of hertz~\cite{Bernard1997, Acef1999} --- by stabilizing these lasers to saturated absorption lines of molecules such as OsO$_4$. CO$_2$ lasers have a major shortcoming: a lack of tunability. They emit at CO$_2$ molecular resonances. An emission line is found every 30 to 50~GHz in the 9-11~$\mu$m wavelength range, and each line is tunable over about 100~MHz. Although, as in our spectrometers, this range can be extended a few gigahertz using electro-optical modulators (EOMs), this is done at the expense of power (EOMs at these wavelengths have an efficiency of 10$^{-4}$) and necessitates subsequent spectral filtering. Overcoming these difficulties without the loss of stability is key to enabling precision measurements in the mid-IR.

One solution would be to use frequency comb-referenced continuous-wave (cw)~\cite{Galli2011, Bressel2012, Ricciardi2012} or femtosecond~\cite{Schliesser2012} mid-IR sources. These are based on frequency mixing in nonlinear crystals and provide absolute-frequency referencing, reasonable line widths and tunability, but are very complex and often exhibit limited power. By comparison, cw quantum cascade lasers (QCLs) are a new mature and robust technology that offer broad and continuous tuning over several hundred gigahertz at 100~mW-level powers. Several can be combined giving access to the whole mid-IR region. Recent studies of the emission spectrum of cw free-running distributed-feedback (DFB) QCLs~\cite{Myers2002, Bartalini2010, Tombez2011, Bartalini2011, Mills2012} confirm their suitability for high resolution spectroscopy and frequency metrology. Furthermore, narrow-emission, absolutely referenced mid-IR QCLs have been demonstrated, either by phase-locking to a CO$_2$ laser~\cite{Bielsa2007, Bielsa2008}, frequency locking to a sub-Doppler molecular transition~\cite{Cappelli2012}, optical injection locking~\cite{Borri2012} or phase-locking~\cite{Mills2012, Hansen2013, Galli2013} to narrow optical frequency comb-based sources. Sub-kHz emission line widths, corresponding to relative stabilities in the high $10^{-13}$, and accuracies of a few $10^{-12}$ have been shown~\cite{Galli2013}. Note however that most of this work has been done using QCLs emitting around 4-$5~\mu$m. Work at longer wavelengths (including most of the molecular fingerprint region) has remained scarce (see refs.~\cite{Bielsa2007, Bielsa2008, Mills2012} around $9~\mu$m).

We extended this range to 10~$\mu$m. A 10~$\mu$m QCL exhibiting remarkably low free-running frequency noise is coherently phase-locked to the OsO$_4$-stabilized CO$_2$ laser. This allows both line width narrowing, by about 4 orders of magnitude down to an unprecedented 10~Hz-level, and absolute frequency referencing at the $10^{-12}$ level. In order to preserve some of its tunability, the QCL is in fact locked to one of two optical sidebands (tunable over 10~GHz) generated by coupling the CO$_2$ laser light through an EOM. Once locked, we use this QCL to demonstrate high-resolution spectroscopy of both NH$_3$ and methyltrioxorhenium (MTO) over a range of over 6~GHz. These species are of interest for the two precision measurements under progress in our group. The former is our molecule of choice for measuring k$_{B}$, while the latter, MTO is an achiral test organometallic complex whose chiral derivatives are considered for a parity violation test~\cite{Darquie2010, Stoeffler2011, Tokunaga2013}.

\begin{figure}
\includegraphics[width=\columnwidth]{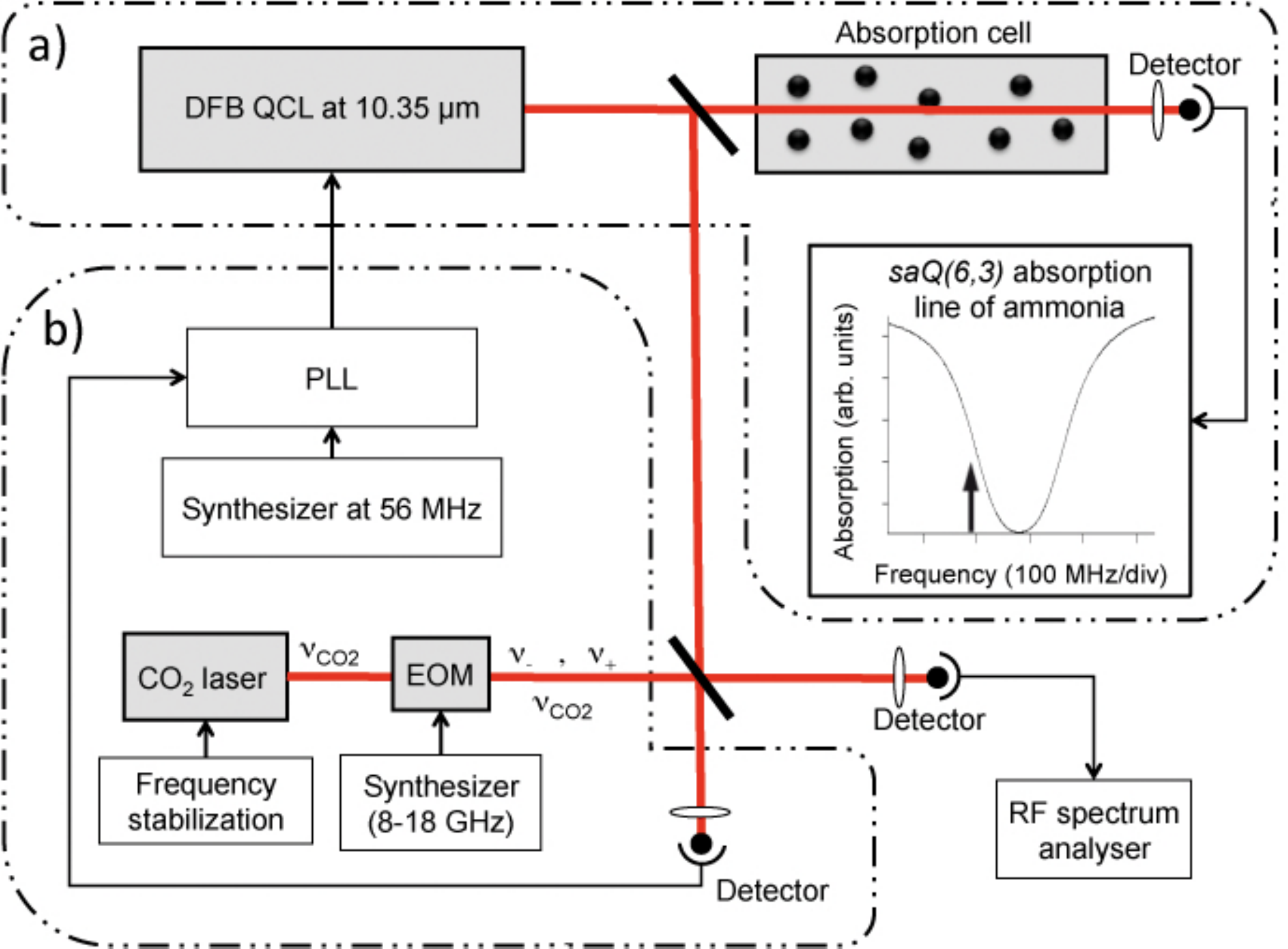}
\caption{Sketch of the experimental setup.(a) Two different absorption of 5~cm and 60~cm-path length cells are respectively used for the free-running QCL frequency noise analysis and for high resolution spectroscopy. The graph shows the measured $saQ(6,3)$ rovibrational line of the $\nu_2$ vibrational mode of $^{14}$NH$_3$ used as frequency discriminator (250~Pa) recorded with our ultra-stable CO$_2$ laser based spectrometer~\cite{Lemarchand2010}. The abscissa shows the detuning from the line center (28~953~693.9(1)~MHz). The arrow represents the QCL operating point when recording the frequency noise power spectral density shown in Fig. \ref{PSD}. (b) Setup used to coherently phase-lock the QCL to a frequency-stabilized CO$_2$ laser. QCL: quantum cascade laser, DFB: distributed-feedback, EOM: electro-optic modulator.\label{setup}}
\end{figure}

\begin{figure}
\includegraphics[width=\columnwidth]{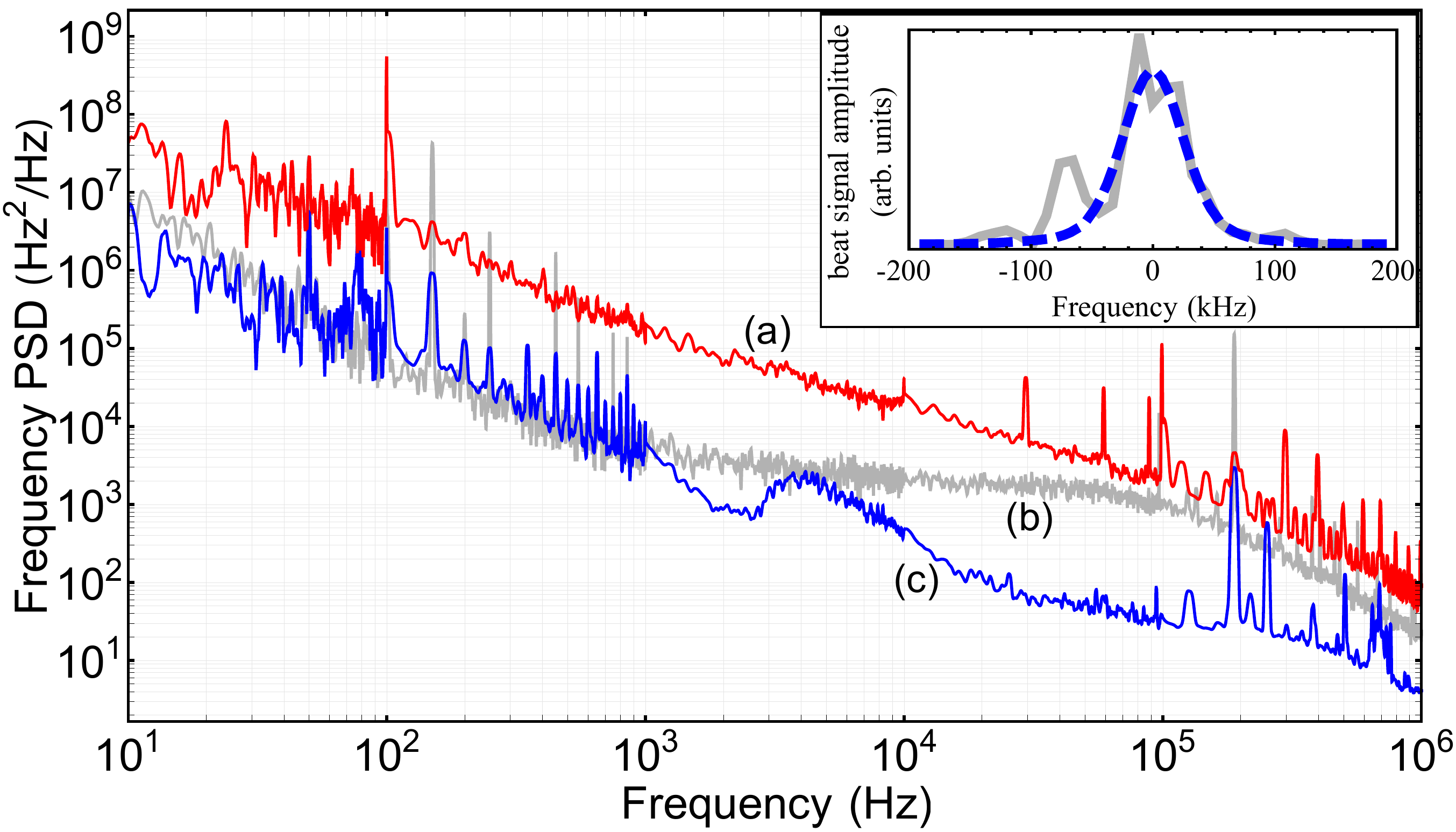}
\caption{Frequency noise PSD of the 10.35~$\mu$m DFB QCL (red line (a)). The contributions of the laser intensity noise (blue line (c)) and laser driver current noise (green line (b)) are also plotted for comparison. The inset shows the beat signal between the QCL and the sideband of a $\sim1$~kHz wide free-running CO$_2$ laser recorded with a RF spectrum analyzer (1~kHz resolution bandwidth). The dashed line is the QCL line shape calculated from the measured frequency noise PSD. For these measurements the QCL was conveniently locked to the side of the $saQ(6,3)$ NH$_3$ line with a $\sim1.5$~Hz bandwidth.\label{PSD}}
\end{figure}

The QCL is a cw-mode near-room-temperature (near-RT) single-mode DFB laser (from Alpes Lasers) tunable between 10.34 and 10.42~$\mu$m (28.76 to 29.00~THz). For the experiments described in this paper, it is typically operated at a temperature of 243~K (at which the threshold current is 710~mA) and a current ranging from 0.96 to 1.02~A, delivering 40 to 60~mW around 28.95~THz. At the QCL output, the laser beam is collimated with a spherical ZnSe lens ($f=25$~mm)
First, we measure the frequency noise of the free-running QCL. This allows us to estimate the feedback bandwidth required to narrow its line width. As illustrated in Fig.~\ref{setup}(a), we use the side of an ammonia linear absorption line as frequency discriminator. The absorption signal from a 5~cm path length cell containing 250~Pa of NH$_3$ is recorded with a liquid nitrogen-cooled HgCdTe detector (with a bandwidth of a few megahertz) and processed by a Fast Fourier Transform (FFT) spectrum analyzer. The frequency-to-amplitude conversion coefficient is measured by recording the same rovibrational line using our stabilized CO$_2$ laser spectrometer~\cite{Lemarchand2010}. Fig.~\ref{PSD} shows the resulting frequency noise power spectral density (PSD) of the QCL. It has a $1/f$ trend at low frequency, followed by a steeper slope above $\sim300$~kHz, as observed in~\cite{Bartalini2010, Tombez2011}. Note however that the measured frequency noise PSD is roughly one order of magnitude lower than previously published characterizations of free-running cw-mode near-RT DFB QCLs~\cite{Myers2002, Bartalini2010, Tombez2011, Bartalini2011, Mills2012}. Fig.~\ref{PSD} also shows the contribution from the laser intensity noise, obtained with the laser tuned far off resonance, as well as the contribution from a home-made low-noise current source. The latter is obtained by multiplying the driver's current noise spectrum ($<300~\mathrm{pA}/\sqrt{\mathrm{Hz}}$ above 10~kHz) by the laser DC current-to-frequency response (230~MHz/mA). The driver's current noise was accurately measured by balancing two identical home-made sources of opposite polarity and detecting the residual AC-currents.

Fig.~\ref{PSD} also shows that the expected white noise level $N_{\mathrm{w}}$ corresponding to the Schawlow-Townes limit does not seem to be reached at 1~MHz. Thus, only an upper limit of $N_{\mathrm{w}}\sim50$~Hz$^2$/Hz (and an upper limit of the corresponding intrinsic laser line width  of $\Delta\nu=\pi N_{\mathrm{w}}\sim160$~Hz) can be inferred. This is 1.7 times lower than measured for a cw mode near-RT DFB QCL at 4.3~$\mu$m~\cite{Bartalini2011}.

The real laser line width is broadened by flicker noise and depends on the observation time. The inset in Fig.~\ref{PSD} shows a beat signal between the free-running QCL and a free-running CO$_2$ laser which exhibits a record $\sim60$~kHz full width at half maximum after 1~ms of integration time (\textit{i.e.} 1~kHz resolution bandwidth). The line width of the free-running CO$_2$ laser was measured to be $\sim1$~kHz~\cite{Bernard1995}, making its contribution negligible in the observed width. The measured beat signal agrees well  with a theoretical estimation of the QCL emission line shape based on the measured frequency noise PSD (following~\cite{Elliott1982} and accounting for the 1~ms observation time~\cite{Bishof2013}) as indicated by the dashed line in the inset of Fig.~\ref{PSD}.

The experimental setup used to coherently phase-lock the QCL to a CO$_2$ laser stabilized on a saturated absorption line of OsO$_4$ is shown on Fig.~\ref{setup}(b). In this work, the $R(6)$ CO$_2$ laser carrier is frequency-locked to the $R(23)A^1_1(-)$ line of $^{192}$OsO$_4$. Once stabilized, light from the CO$_2$ laser is coupled to an EOM which generates two sidebands of frequencies $\nu_{\pm}=\nu_{\mathrm{CO_2}}\pm f_{\mathrm{EOM}}$ on either side of the fixed laser frequency $\nu_{\mathrm{CO_2}}$. The frequency $f_{\mathrm{EOM}}$ is tunable from 8 to 18~GHz. The QCL ($\sim100~\mu$W) and the CO$_2$ laser beams are overlapped and the beat signal is detected by a liquid nitrogen-cooled HgCdTe detector with a bandwidth of about 100~MHz. With a $\sim10^{-4}$ EOM efficiency, $\sim1~\mu$W ($\sim10~$mW) is available for the beat signal in each of the CO$_2$ laser sidebands (in the carrier). 

The phase-error signal is generated by comparing the phase of the amplified (60~dB) beat signal with a synthesized reference signal typically at 56~MHz using a frequency mixer. A phase-lock servo loop is used to apply a correction signal directly to the QCL's current.

\begin{figure}
\includegraphics[width=\columnwidth]{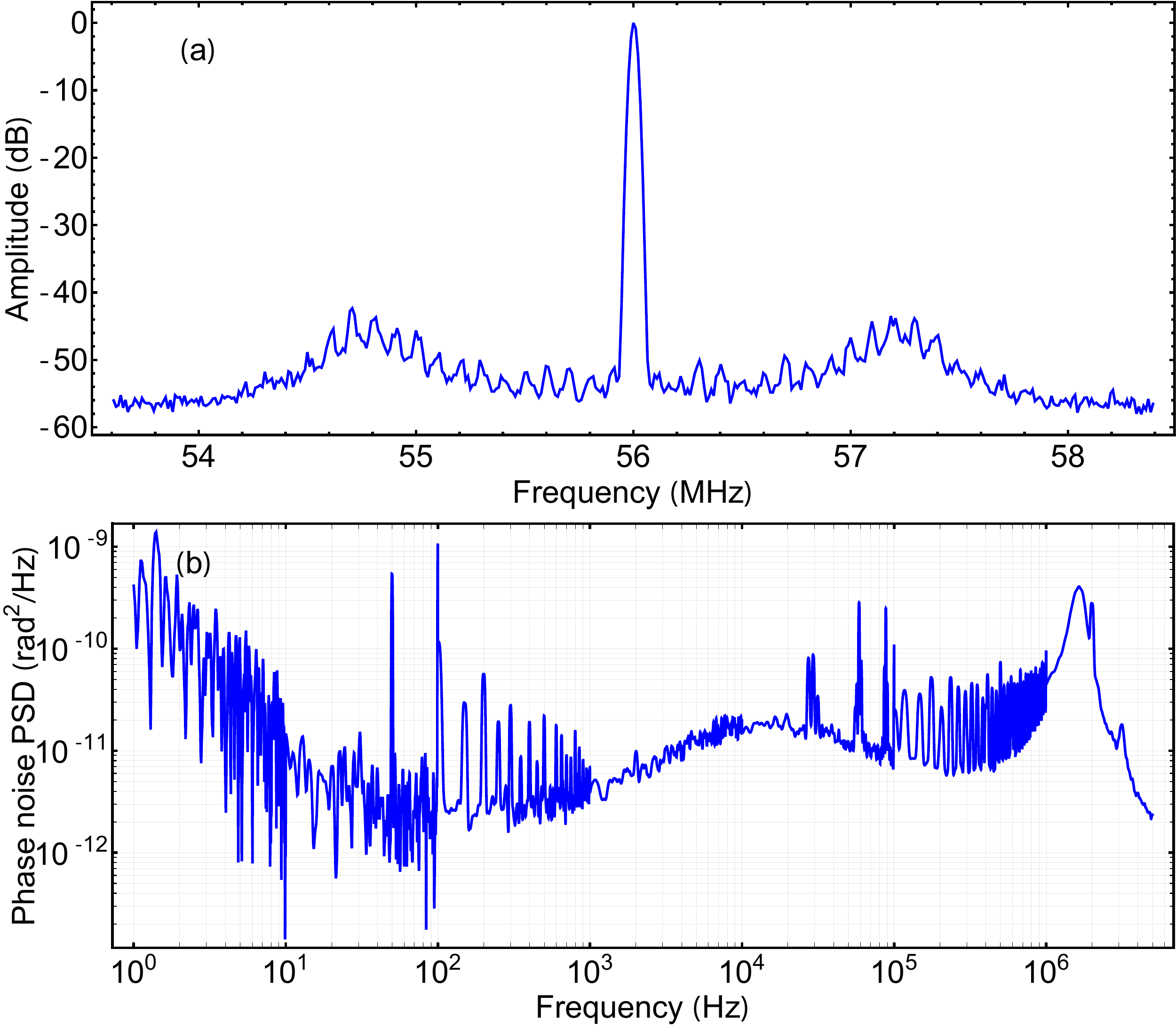}
\caption{(a) Beat signal spectrum between the frequency-stabilized CO$_2$ laser sideband and the phase-locked QCL taken with a RF spectrum analyzer (30~kHz resolution bandwidth). (b) Phase-noise power spectral density of the beat signal between the QCL and the frequency-stabilized CO$_2$ laser carrier.\label{phase-lock}}
\end{figure}

The beat spectrum between a CO$_2$ laser sideband and the QCL phase-locked to this sideband, processed by a radio-frequency (RF) spectrum analyzer, is shown in Fig.~\ref{phase-lock}(a). It is recorded with a second out-of-loop photodetector (see Fig.~\ref{setup}), similar to the one in-loop, in order to avoid errors brought by the detection setup and associated electronics. It represents the relative phase noise spectral density between the QCL and the CO$_2$ laser. With a typical signal-to-noise ratio of 60~dB in a 30~kHz resolution bandwidth we achieve a feedback bandwidth between 1 and 3~MHz (as indicated by the servo-loop unity gain frequency bump in the spectrum wings). As expected for a proper phase-lock, the beat signal of Fig.~\ref{phase-lock}(a) exhibits an extremely narrow central peak. Its width was observed to be limited by the 10-Hz resolution of our RF spectrum analyzer. 
Fig.~\ref{phase-lock}(b) shows the phase-noise power spectral density of the beat-signal between the CO$_2$ laser carrier and the phase-locked QCL (the beat signal of the out-of-loop photodetector is down-mixed to DC and processed with a FFT spectrum analyzer). The phase-lock performance can be characterized by the residual rms phase error, $\sigma_{\varphi}$. Integration of the phase-noise data in Fig.~\ref{phase-lock}(b) from 1~Hz up to 5~MHz leads to a very small $\sigma_{\varphi}\sim0.017$~rad, corresponding to having $e^{-\sigma_{\varphi}^2}\sim99.97\%$ of the beat signal RF power concentrated in the coherent part (\textit{i.e.} in the central peak)~\cite{Zhu1993}.

This is a signature of a highly coherent phase-lock and of the excellent transfer of the locked CO$_2$ laser's spectral features to the QCL. This results in a record QCL line width of the order of 10~Hz, 3 to 4 orders of magnitude lower than a free-running QCL, and a relative stability at 1~s of about 1~Hz. Our particular choice of OsO$_4$ line for absolute frequency referencing leads to a 90~Hz accuracy of the frequency scale~\cite{Chardonnet1989, Acef1999}. To our knowledge, this is the first demonstration that a QCL can reach these metrological spectral properties.

\begin{figure}
\includegraphics[width=\columnwidth]{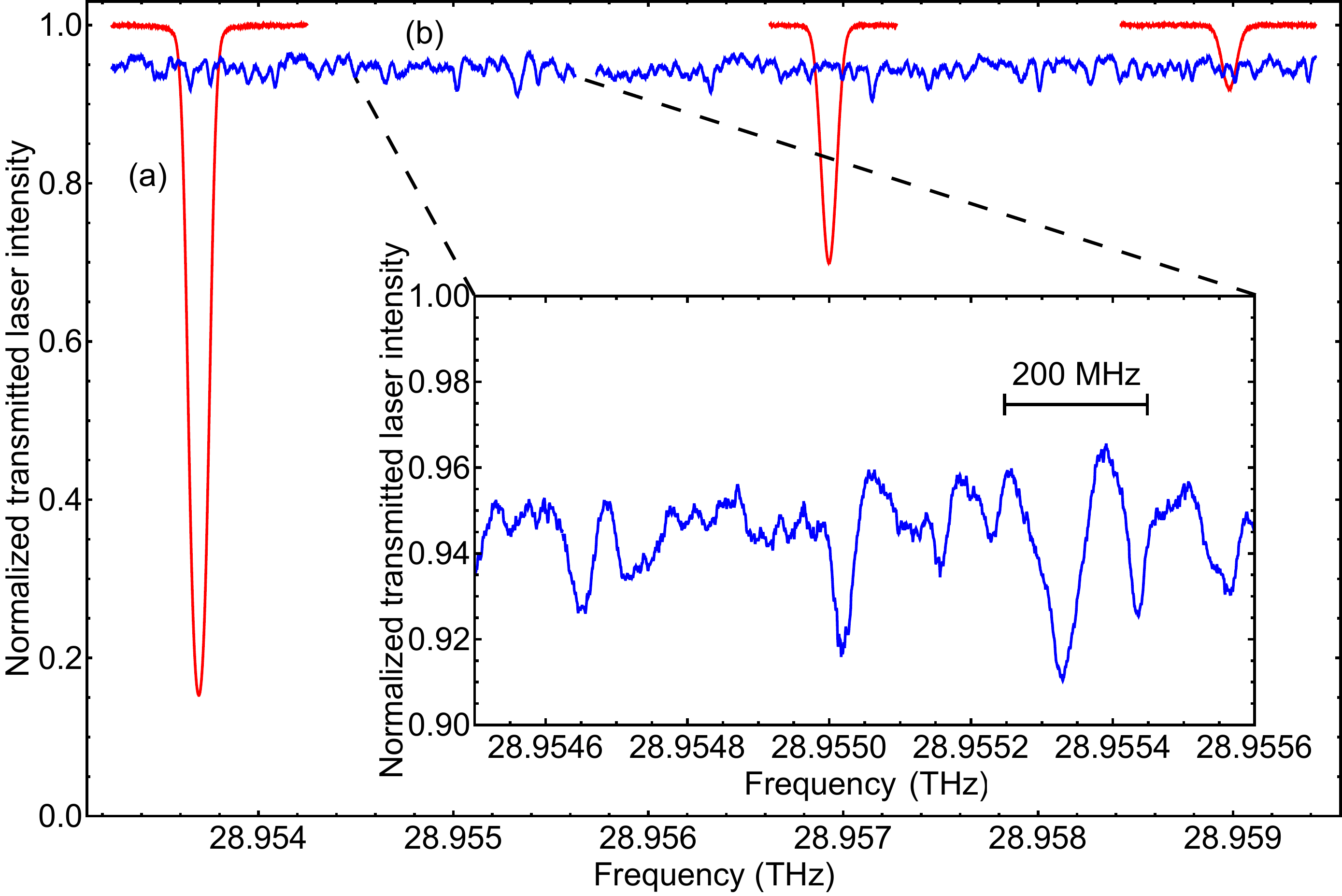}
\caption{Linear absorption spectra of NH$_3$ (red curve (a)) and MTO (blue curve (b)) recorded over more than 6~GHz with a $\sim10~$Hz line width QCL phase-locked to a frequency-stabilized CO$_2$ laser. From left to right, the ammonia spectrum exhibit the three isolated rovibrational lines $saQ(6,3)$ (probed in our experiment dedicated to the determination of $k_{\mathrm{B}}$~\cite{Lemarchand2013}), $saQ(6,2)$ and $saQ(6,1)$ of the $\nu_2$ vibrational mode of $^{14}$NH$_3$. The inset shows a zoom on the MTO spectrum. Experimental conditions: chopper frequency 2~kHz, lock-in amplifier time constant 30~ms, 500~kHz steps, power $1~\mu$W, absorption length 60 cm, $\sim10$~Pa of NH$_3$ or $\sim4$~Pa of MTO. \label{spectra}}
\end{figure}

Linear absorption spectra of both NH$_3$ and MTO are shown on Fig.~\ref{spectra}. The setup has been described above, and is depicted in Fig.~\ref{setup}. A few-millimeter wide beam of $1~\mu$W is sent through a 60-cm long cell filled with either $\sim10$~Pa of ammonia or $\sim4$~Pa of MTO. A slotted disk, that chops the beam at 2~kHz, and a lock-in detection are used for noise filtering. The QCL is phase-locked to the EOM's negative sideband (frequency $\nu_-$), red-detuned from the CO$_2$ laser carrier (frequency $\nu_\mathrm{CO_2}$).  Sweeping the EOM frequency enables us to continuously tune the QCL over $\sim6$~GHz, more than $10^8$ times the laser line width. With a few milliwatts available for spectroscopy, this also results in an effective power amplification of $\sim10^3$ compared to using the CO$_2$ laser beam's negative sideband directly. The ammonia spectrum in Fig.~\ref{spectra} exhibits three isolated rovibrational lines of the $\nu_2$ vibrational mode of $^{14}$NH$_3$. The MTO spectrum looks very different with a mean $\sim5.5\%$ absorption and $\sim100~$MHz wide $\sim2\%$ deviations as expected from the dense antisymmetric Re=O stretching mode of the molecule~\cite{Darquie2010, Stoeffler2011, Tokunaga2013}. Both spectra were normalized by numerically correcting for the baseline.

In conclusion we characterized the frequency noise of a free-running cw near-RT DFB $10.3~\mu$m QCL. The laser is then phase-locked to a frequency-stabilized CO$_2$ laser, and we observe that the spectral properties of the latter are successfully copied to the QCL. This results in a record line width of the order of 10~Hz, a relative stability at 1~s in the $10^{-14}$ range and a relative accuracy of $3\times10^{-12}$. Spectra of ammonia and MTO over several GHz using our QCL source were presented, thereby demonstrating the potential of QCLs for precision measurements devoted to metrological applications or tests of fundamental laws of nature.

The use of QCLs will eventually allow the study of any species showing absorption between 3 and 25~$\mu$m, with much broader continuous tuning range and with reasonable powers. It will broaden the scope of our spectroscopic precision measurement experiments using molecules.
We eventually plan to reach the $3\times10^{-16}$ accuracy of the Cs fountain~\cite{Guena2012}, the $10^{-15}$ stability of the best near-IR oscillators, and take full advantage of the QCL's tunability by locking it to a frequency comb stabilized via on optical fiber to an ultra-stable near-IR reference monitored against atomic fountain clocks. This was recently demonstrated in our group with a CO$_2$ lasers~\cite{Chanteau2013}. Stabilizing the laser this way provides the ultimate frequency accuracy and stability, and frees us from having to lock the QCL to any particular reference (another laser or a molecular transition), which would constrain the laser's operating frequency. Finally, narrow line width light sources such as QCLs would benefit the entire spectroscopy community, beyond our scope of interest. Commercial QCLs are available at wavelengths spanning the entire mid-IR, the molecular fingerprint region, which hosts many spectral signatures of molecules of interest for atmospheric, planetary or interstellar physics, chemistry, biology, medical or industrial diagnostics.


The authors acknowledge financial support from CNRS, Universit\'{e} Paris 13, LNE, LabEx FIRST-TF (ANR-IO-LABX-48-01) and AS GRAM. This work is part of the projects NCPCHEM n\textsuperscript{o} 2010 BLAN 724 3 and QUIGARDE n\textsuperscript{o} ANR-12-ASTR-0028-03 funded by the Agence Nationale de la Recherche (ANR, France).


\nocite{*}
\bibliography{QCLbib}

\begin{thebibliography}{39}%
\makeatletter
\providecommand \@ifxundefined [1]{%
 \@ifx{#1\undefined}
}%
\providecommand \@ifnum [1]{%
 \ifnum #1\expandafter \@firstoftwo
 \else \expandafter \@secondoftwo
 \fi
}%
\providecommand \@ifx [1]{%
 \ifx #1\expandafter \@firstoftwo
 \else \expandafter \@secondoftwo
 \fi
}%
\providecommand \natexlab [1]{#1}%
\providecommand \enquote  [1]{``#1''}%
\providecommand \bibnamefont  [1]{#1}%
\providecommand \bibfnamefont [1]{#1}%
\providecommand \citenamefont [1]{#1}%
\providecommand \href@noop [0]{\@secondoftwo}%
\providecommand \href [0]{\begingroup \@sanitize@url \@href}%
\providecommand \@href[1]{\@@startlink{#1}\@@href}%
\providecommand \@@href[1]{\endgroup#1\@@endlink}%
\providecommand \@sanitize@url [0]{\catcode `\\12\catcode `\$12\catcode
  `\&12\catcode `\#12\catcode `\^12\catcode `\_12\catcode `\%12\relax}%
\providecommand \@@startlink[1]{}%
\providecommand \@@endlink[0]{}%
\providecommand \url  [0]{\begingroup\@sanitize@url \@url }%
\providecommand \@url [1]{\endgroup\@href {#1}{\urlprefix }}%
\providecommand \urlprefix  [0]{URL }%
\providecommand \Eprint [0]{\href }%
\providecommand \doibase [0]{http://dx.doi.org/}%
\providecommand \selectlanguage [0]{\@gobble}%
\providecommand \bibinfo  [0]{\@secondoftwo}%
\providecommand \bibfield  [0]{\@secondoftwo}%
\providecommand \translation [1]{[#1]}%
\providecommand \BibitemOpen [0]{}%
\providecommand \bibitemStop [0]{}%
\providecommand \bibitemNoStop [0]{.\EOS\space}%
\providecommand \EOS [0]{\spacefactor3000\relax}%
\providecommand \BibitemShut  [1]{\csname bibitem#1\endcsname}%
\let\auto@bib@innerbib\@empty
\bibitem [{\citenamefont {Daussy}\ \emph {et~al.}(1999)\citenamefont {Daussy},
  \citenamefont {Marrel}, \citenamefont {Amy-Klein}, \citenamefont {Nguyen},
  \citenamefont {Bord\'{e}},\ and\ \citenamefont {Chardonnet}}]{Daussy1999}%
  \BibitemOpen
  \bibfield  {author} {\bibinfo {author} {\bibfnamefont {C.}~\bibnamefont
  {Daussy}}, \bibinfo {author} {\bibfnamefont {T.}~\bibnamefont {Marrel}},
  \bibinfo {author} {\bibfnamefont {A.}~\bibnamefont {Amy-Klein}}, \bibinfo
  {author} {\bibfnamefont {C.~T.}\ \bibnamefont {Nguyen}}, \bibinfo {author}
  {\bibfnamefont {C.}~\bibnamefont {Bord\'{e}}}, \ and\ \bibinfo {author}
  {\bibfnamefont {C.}~\bibnamefont {Chardonnet}},\ }\href@noop {} {\bibfield
  {journal} {\bibinfo  {journal} {Phys. Rev. Lett.}\ }\textbf {\bibinfo
  {volume} {83}},\ \bibinfo {pages} {1554} (\bibinfo {year}
  {1999})}\BibitemShut {NoStop}%
\bibitem [{\citenamefont {DeMille}\ \emph {et~al.}(2008)\citenamefont
  {DeMille}, \citenamefont {Cahn}, \citenamefont {Murphree}, \citenamefont
  {Rahmlow},\ and\ \citenamefont {Kozlov}}]{DeMille2008}%
  \BibitemOpen
  \bibfield  {author} {\bibinfo {author} {\bibfnamefont {D.}~\bibnamefont
  {DeMille}}, \bibinfo {author} {\bibfnamefont {S.~B.}\ \bibnamefont {Cahn}},
  \bibinfo {author} {\bibfnamefont {D.}~\bibnamefont {Murphree}}, \bibinfo
  {author} {\bibfnamefont {D.~A.}\ \bibnamefont {Rahmlow}}, \ and\ \bibinfo
  {author} {\bibfnamefont {M.~G.}\ \bibnamefont {Kozlov}},\ }\href {<Go to
  ISI>://18232864} {\bibfield  {journal} {\bibinfo  {journal} {Phys. Rev.
  Lett.}\ }\textbf {\bibinfo {volume} {100}},\ \bibinfo {pages} {23003}
  (\bibinfo {year} {2008})}\BibitemShut {NoStop}%
\bibitem [{\citenamefont {Darqui\'{e}}\ \emph {et~al.}(2010)\citenamefont
  {Darqui\'{e}}, \citenamefont {Stoeffler}, \citenamefont {Shelkovnikov},
  \citenamefont {Daussy}, \citenamefont {Amy-Klein}, \citenamefont
  {Chardonnet}, \citenamefont {Zrig}, \citenamefont {Guy}, \citenamefont
  {Crassous}, \citenamefont {Soulard}, \citenamefont {Asselin}, \citenamefont
  {Huet}, \citenamefont {Schwerdtfeger}, \citenamefont {Bast},\ and\
  \citenamefont {Saue}}]{Darquie2010}%
  \BibitemOpen
  \bibfield  {author} {\bibinfo {author} {\bibfnamefont {B.}~\bibnamefont
  {Darqui\'{e}}}, \bibinfo {author} {\bibfnamefont {C.}~\bibnamefont
  {Stoeffler}}, \bibinfo {author} {\bibfnamefont {A.}~\bibnamefont
  {Shelkovnikov}}, \bibinfo {author} {\bibfnamefont {C.}~\bibnamefont
  {Daussy}}, \bibinfo {author} {\bibfnamefont {A.}~\bibnamefont {Amy-Klein}},
  \bibinfo {author} {\bibfnamefont {C.}~\bibnamefont {Chardonnet}}, \bibinfo
  {author} {\bibfnamefont {S.}~\bibnamefont {Zrig}}, \bibinfo {author}
  {\bibfnamefont {L.}~\bibnamefont {Guy}}, \bibinfo {author} {\bibfnamefont
  {J.}~\bibnamefont {Crassous}}, \bibinfo {author} {\bibfnamefont
  {P.}~\bibnamefont {Soulard}}, \bibinfo {author} {\bibfnamefont
  {P.}~\bibnamefont {Asselin}}, \bibinfo {author} {\bibfnamefont {T.~R.}\
  \bibnamefont {Huet}}, \bibinfo {author} {\bibfnamefont {P.}~\bibnamefont
  {Schwerdtfeger}}, \bibinfo {author} {\bibfnamefont {R.}~\bibnamefont {Bast}},
  \ and\ \bibinfo {author} {\bibfnamefont {T.}~\bibnamefont {Saue}},\ }\href
  {<Go to ISI>://000282652300003} {\bibfield  {journal} {\bibinfo  {journal}
  {Chirality}\ }\textbf {\bibinfo {volume} {22}},\ \bibinfo {pages} {870}
  (\bibinfo {year} {2010})}\BibitemShut {NoStop}%
\bibitem [{\citenamefont {Baron}\ \emph {et~al.}(2014)\citenamefont {Baron},
  \citenamefont {Campbell}, \citenamefont {DeMille}, \citenamefont {Doyle},
  \citenamefont {Gabrielse}, \citenamefont {Gurevich}, \citenamefont {Hess},
  \citenamefont {Hutzler}, \citenamefont {Kirilov}, \citenamefont {Kozyryev},
  \citenamefont {O'Leary}, \citenamefont {Panda}, \citenamefont {Parsons},
  \citenamefont {Petrik}, \citenamefont {Spaun}, \citenamefont {Vutha},\ and\
  \citenamefont {West}}]{Baron2014}%
  \BibitemOpen
  \bibfield  {author} {\bibinfo {author} {\bibfnamefont {J.}~\bibnamefont
  {Baron}}, \bibinfo {author} {\bibfnamefont {W.~C.}\ \bibnamefont {Campbell}},
  \bibinfo {author} {\bibfnamefont {D.}~\bibnamefont {DeMille}}, \bibinfo
  {author} {\bibfnamefont {J.~M.}\ \bibnamefont {Doyle}}, \bibinfo {author}
  {\bibfnamefont {G.}~\bibnamefont {Gabrielse}}, \bibinfo {author}
  {\bibfnamefont {Y.~V.}\ \bibnamefont {Gurevich}}, \bibinfo {author}
  {\bibfnamefont {P.~W.}\ \bibnamefont {Hess}}, \bibinfo {author}
  {\bibfnamefont {N.~R.}\ \bibnamefont {Hutzler}}, \bibinfo {author}
  {\bibfnamefont {E.}~\bibnamefont {Kirilov}}, \bibinfo {author} {\bibfnamefont
  {I.}~\bibnamefont {Kozyryev}}, \bibinfo {author} {\bibfnamefont {B.~R.}\
  \bibnamefont {O'Leary}}, \bibinfo {author} {\bibfnamefont {C.~D.}\
  \bibnamefont {Panda}}, \bibinfo {author} {\bibfnamefont {M.~F.}\ \bibnamefont
  {Parsons}}, \bibinfo {author} {\bibfnamefont {E.~S.}\ \bibnamefont {Petrik}},
  \bibinfo {author} {\bibfnamefont {B.}~\bibnamefont {Spaun}}, \bibinfo
  {author} {\bibfnamefont {A.~C.}\ \bibnamefont {Vutha}}, \ and\ \bibinfo
  {author} {\bibfnamefont {A.~D.}\ \bibnamefont {West}},\ }\href {\doibase
  10.1126/science.1248213} {\bibfield  {journal} {\bibinfo  {journal}
  {Science}\ }\textbf {\bibinfo {volume} {343}},\ \bibinfo {pages} {269}
  (\bibinfo {year} {2014})}\BibitemShut {NoStop}%
\bibitem [{\citenamefont {Koelemeij}\ \emph {et~al.}(2007)\citenamefont
  {Koelemeij}, \citenamefont {Roth}, \citenamefont {Wicht}, \citenamefont
  {Ernsting},\ and\ \citenamefont {Schiller}}]{Koelemeij2007}%
  \BibitemOpen
  \bibfield  {author} {\bibinfo {author} {\bibfnamefont {J.~C.~J.}\
  \bibnamefont {Koelemeij}}, \bibinfo {author} {\bibfnamefont {B.}~\bibnamefont
  {Roth}}, \bibinfo {author} {\bibfnamefont {A.}~\bibnamefont {Wicht}},
  \bibinfo {author} {\bibfnamefont {I.}~\bibnamefont {Ernsting}}, \ and\
  \bibinfo {author} {\bibfnamefont {S.}~\bibnamefont {Schiller}},\ }\href
  {http://link.aps.org/doi/10.1103/PhysRevLett.98.173002} {\bibfield  {journal}
  {\bibinfo  {journal} {Phys. Rev. Lett.}\ }\textbf {\bibinfo {volume} {98}},\
  \bibinfo {pages} {173002} (\bibinfo {year} {2007})}\BibitemShut {NoStop}%
\bibitem [{\citenamefont {Lemarchand}\ \emph {et~al.}(2013)\citenamefont
  {Lemarchand}, \citenamefont {Mejri}, \citenamefont {Sow}, \citenamefont
  {Triki}, \citenamefont {Tokunaga}, \citenamefont {Briaudeau}, \citenamefont
  {Chardonnet}, \citenamefont {Darqui\'{e}},\ and\ \citenamefont
  {Daussy}}]{Lemarchand2013}%
  \BibitemOpen
  \bibfield  {author} {\bibinfo {author} {\bibfnamefont {C.}~\bibnamefont
  {Lemarchand}}, \bibinfo {author} {\bibfnamefont {S.}~\bibnamefont {Mejri}},
  \bibinfo {author} {\bibfnamefont {P.~L.~T.}\ \bibnamefont {Sow}}, \bibinfo
  {author} {\bibfnamefont {M.}~\bibnamefont {Triki}}, \bibinfo {author}
  {\bibfnamefont {S.~K.}\ \bibnamefont {Tokunaga}}, \bibinfo {author}
  {\bibfnamefont {S.}~\bibnamefont {Briaudeau}}, \bibinfo {author}
  {\bibfnamefont {C.}~\bibnamefont {Chardonnet}}, \bibinfo {author}
  {\bibfnamefont {B.}~\bibnamefont {Darqui\'{e}}}, \ and\ \bibinfo {author}
  {\bibfnamefont {C.}~\bibnamefont {Daussy}},\ }\href {\doibase
  10.1088/0026-1394/50/6/623} {\bibfield  {journal} {\bibinfo  {journal}
  {Metrologia}\ }\textbf {\bibinfo {volume} {50}},\ \bibinfo {pages} {623}
  (\bibinfo {year} {2013})}\BibitemShut {NoStop}%
\bibitem [{\citenamefont {Moretti}\ \emph {et~al.}(2013)\citenamefont
  {Moretti}, \citenamefont {Castrillo}, \citenamefont {Fasci}, \citenamefont
  {{De Vizia}}, \citenamefont {Casa}, \citenamefont {Galzerano}, \citenamefont
  {Merlone}, \citenamefont {Laporta},\ and\ \citenamefont
  {Gianfrani}}]{Moretti2013}%
  \BibitemOpen
  \bibfield  {author} {\bibinfo {author} {\bibfnamefont {L.}~\bibnamefont
  {Moretti}}, \bibinfo {author} {\bibfnamefont {A.}~\bibnamefont {Castrillo}},
  \bibinfo {author} {\bibfnamefont {E.}~\bibnamefont {Fasci}}, \bibinfo
  {author} {\bibfnamefont {M.~D.}\ \bibnamefont {{De Vizia}}}, \bibinfo
  {author} {\bibfnamefont {G.}~\bibnamefont {Casa}}, \bibinfo {author}
  {\bibfnamefont {G.}~\bibnamefont {Galzerano}}, \bibinfo {author}
  {\bibfnamefont {A.}~\bibnamefont {Merlone}}, \bibinfo {author} {\bibfnamefont
  {P.}~\bibnamefont {Laporta}}, \ and\ \bibinfo {author} {\bibfnamefont
  {L.}~\bibnamefont {Gianfrani}},\ }\href {\doibase
  10.1103/PhysRevLett.111.060803} {\bibfield  {journal} {\bibinfo  {journal}
  {Physical Review Letters}\ }\textbf {\bibinfo {volume} {111}},\ \bibinfo
  {pages} {060803} (\bibinfo {year} {2013})}\BibitemShut {NoStop}%
\bibitem [{\citenamefont {Hudson}\ \emph {et~al.}(2006)\citenamefont {Hudson},
  \citenamefont {Lewandowski}, \citenamefont {Sawyer},\ and\ \citenamefont
  {Ye}}]{Hudson2006}%
  \BibitemOpen
  \bibfield  {author} {\bibinfo {author} {\bibfnamefont {E.~R.}\ \bibnamefont
  {Hudson}}, \bibinfo {author} {\bibfnamefont {H.~J.}\ \bibnamefont
  {Lewandowski}}, \bibinfo {author} {\bibfnamefont {B.~C.}\ \bibnamefont
  {Sawyer}}, \ and\ \bibinfo {author} {\bibfnamefont {J.}~\bibnamefont {Ye}},\
  }\href {http://link.aps.org/doi/10.1103/PhysRevLett.96.143004} {\bibfield
  {journal} {\bibinfo  {journal} {Phys. Rev. Lett.}\ }\textbf {\bibinfo
  {volume} {96}},\ \bibinfo {pages} {143004} (\bibinfo {year}
  {2006})}\BibitemShut {NoStop}%
\bibitem [{\citenamefont {Shelkovnikov}\ \emph {et~al.}(2008)\citenamefont
  {Shelkovnikov}, \citenamefont {Butcher}, \citenamefont {Chardonnet},\ and\
  \citenamefont {Amy-Klein}}]{Shelkovnikov2008}%
  \BibitemOpen
  \bibfield  {author} {\bibinfo {author} {\bibfnamefont {A.}~\bibnamefont
  {Shelkovnikov}}, \bibinfo {author} {\bibfnamefont {R.~J.}\ \bibnamefont
  {Butcher}}, \bibinfo {author} {\bibfnamefont {C.}~\bibnamefont {Chardonnet}},
  \ and\ \bibinfo {author} {\bibfnamefont {A.}~\bibnamefont {Amy-Klein}},\
  }\href@noop {} {\bibfield  {journal} {\bibinfo  {journal} {Phys. Rev. Lett.}\
  }\textbf {\bibinfo {volume} {100}},\ \bibinfo {pages} {150801} (\bibinfo
  {year} {2008})}\BibitemShut {NoStop}%
\bibitem [{\citenamefont {Truppe}\ \emph {et~al.}(2013)\citenamefont {Truppe},
  \citenamefont {Hendricks}, \citenamefont {Tokunaga}, \citenamefont
  {Lewandowski}, \citenamefont {Kozlov}, \citenamefont {Henkel}, \citenamefont
  {Hinds},\ and\ \citenamefont {Tarbutt}}]{Truppe2013}%
  \BibitemOpen
  \bibfield  {author} {\bibinfo {author} {\bibfnamefont {S.}~\bibnamefont
  {Truppe}}, \bibinfo {author} {\bibfnamefont {R.~J.}\ \bibnamefont
  {Hendricks}}, \bibinfo {author} {\bibfnamefont {S.~K.}\ \bibnamefont
  {Tokunaga}}, \bibinfo {author} {\bibfnamefont {H.~J.}\ \bibnamefont
  {Lewandowski}}, \bibinfo {author} {\bibfnamefont {M.~G.}\ \bibnamefont
  {Kozlov}}, \bibinfo {author} {\bibfnamefont {C.}~\bibnamefont {Henkel}},
  \bibinfo {author} {\bibfnamefont {E.~A.}\ \bibnamefont {Hinds}}, \ and\
  \bibinfo {author} {\bibfnamefont {M.~R.}\ \bibnamefont {Tarbutt}},\ }\href
  {\doibase 10.1038/ncomms3600} {\bibfield  {journal} {\bibinfo  {journal}
  {Nature communications}\ }\textbf {\bibinfo {volume} {4}},\ \bibinfo {pages}
  {2600} (\bibinfo {year} {2013})}\BibitemShut {NoStop}%
\bibitem [{\citenamefont {Ziskind}\ \emph {et~al.}(2002)\citenamefont
  {Ziskind}, \citenamefont {Daussy}, \citenamefont {Marrel},\ and\
  \citenamefont {Chardonnet}}]{Ziskind2002}%
  \BibitemOpen
  \bibfield  {author} {\bibinfo {author} {\bibfnamefont {M.}~\bibnamefont
  {Ziskind}}, \bibinfo {author} {\bibfnamefont {C.}~\bibnamefont {Daussy}},
  \bibinfo {author} {\bibfnamefont {T.}~\bibnamefont {Marrel}}, \ and\ \bibinfo
  {author} {\bibfnamefont {C.}~\bibnamefont {Chardonnet}},\ }\href {\doibase
  10.1140/epjd/e2002-00133-0} {\bibfield  {journal} {\bibinfo  {journal} {The
  European Physical Journal D}\ }\textbf {\bibinfo {volume} {20}},\ \bibinfo
  {pages} {219} (\bibinfo {year} {2002})}\BibitemShut {NoStop}%
\bibitem [{\citenamefont {Lemarchand}\ \emph {et~al.}(2011)\citenamefont
  {Lemarchand}, \citenamefont {Triki}, \citenamefont {Darqui\'{e}},
  \citenamefont {Bord\'{e}}, \citenamefont {Chardonnet},\ and\ \citenamefont
  {Daussy}}]{Lemarchand2011}%
  \BibitemOpen
  \bibfield  {author} {\bibinfo {author} {\bibfnamefont {C.}~\bibnamefont
  {Lemarchand}}, \bibinfo {author} {\bibfnamefont {M.}~\bibnamefont {Triki}},
  \bibinfo {author} {\bibfnamefont {B.}~\bibnamefont {Darqui\'{e}}}, \bibinfo
  {author} {\bibfnamefont {C.~J.}\ \bibnamefont {Bord\'{e}}}, \bibinfo {author}
  {\bibfnamefont {C.}~\bibnamefont {Chardonnet}}, \ and\ \bibinfo {author}
  {\bibfnamefont {C.}~\bibnamefont {Daussy}},\ }\href {\doibase
  10.1088/1367-2630/13/7/073028} {\bibfield  {journal} {\bibinfo  {journal}
  {New Journal of Physics}\ }\textbf {\bibinfo {volume} {13}},\ \bibinfo
  {pages} {073028} (\bibinfo {year} {2011})}\BibitemShut {NoStop}%
\bibitem [{\citenamefont {Tokunaga}\ \emph {et~al.}(2013)\citenamefont
  {Tokunaga}, \citenamefont {Stoeffler}, \citenamefont {Auguste}, \citenamefont
  {Shelkovnikov}, \citenamefont {Daussy}, \citenamefont {Amy-Klein},
  \citenamefont {Chardonnet},\ and\ \citenamefont
  {Darqui\'{e}}}]{Tokunaga2013}%
  \BibitemOpen
  \bibfield  {author} {\bibinfo {author} {\bibfnamefont {S.~K.}\ \bibnamefont
  {Tokunaga}}, \bibinfo {author} {\bibfnamefont {C.}~\bibnamefont {Stoeffler}},
  \bibinfo {author} {\bibfnamefont {F.}~\bibnamefont {Auguste}}, \bibinfo
  {author} {\bibfnamefont {A.}~\bibnamefont {Shelkovnikov}}, \bibinfo {author}
  {\bibfnamefont {C.}~\bibnamefont {Daussy}}, \bibinfo {author} {\bibfnamefont
  {A.}~\bibnamefont {Amy-Klein}}, \bibinfo {author} {\bibfnamefont
  {C.}~\bibnamefont {Chardonnet}}, \ and\ \bibinfo {author} {\bibfnamefont
  {B.}~\bibnamefont {Darqui\'{e}}},\ }\href {\doibase
  10.1080/00268976.2013.821186} {\bibfield  {journal} {\bibinfo  {journal}
  {Molecular Physics}\ }\textbf {\bibinfo {volume} {111}},\ \bibinfo {pages}
  {2363} (\bibinfo {year} {2013})}\BibitemShut {NoStop}%
\bibitem [{\citenamefont {Bernard}\ \emph {et~al.}(1997)\citenamefont
  {Bernard}, \citenamefont {Daussy}, \citenamefont {Nogues}, \citenamefont
  {Constantin}, \citenamefont {Durand}, \citenamefont {Amy-Klein},
  \citenamefont {van Lerberghe},\ and\ \citenamefont
  {Chardonnet}}]{Bernard1997}%
  \BibitemOpen
  \bibfield  {author} {\bibinfo {author} {\bibfnamefont {V.}~\bibnamefont
  {Bernard}}, \bibinfo {author} {\bibfnamefont {C.}~\bibnamefont {Daussy}},
  \bibinfo {author} {\bibfnamefont {G.}~\bibnamefont {Nogues}}, \bibinfo
  {author} {\bibfnamefont {L.}~\bibnamefont {Constantin}}, \bibinfo {author}
  {\bibfnamefont {P.~E.}\ \bibnamefont {Durand}}, \bibinfo {author}
  {\bibfnamefont {A.}~\bibnamefont {Amy-Klein}}, \bibinfo {author}
  {\bibfnamefont {A.}~\bibnamefont {van Lerberghe}}, \ and\ \bibinfo {author}
  {\bibfnamefont {C.}~\bibnamefont {Chardonnet}},\ }\href@noop {} {\bibfield
  {journal} {\bibinfo  {journal} {IEEE J. Quant. Elec.}\ }\textbf {\bibinfo
  {volume} {QE-33}},\ \bibinfo {pages} {1282} (\bibinfo {year}
  {1997})}\BibitemShut {NoStop}%
\bibitem [{\citenamefont {Acef}, \citenamefont {Michaud},\ and\ \citenamefont
  {Rovera}(1999)}]{Acef1999}%
  \BibitemOpen
  \bibfield  {author} {\bibinfo {author} {\bibfnamefont {O.}~\bibnamefont
  {Acef}}, \bibinfo {author} {\bibfnamefont {F.}~\bibnamefont {Michaud}}, \
  and\ \bibinfo {author} {\bibfnamefont {G.~V.}\ \bibnamefont {Rovera}},\
  }\href {\doibase 10.1109/19.769659} {\bibfield  {journal} {\bibinfo
  {journal} {IEEE Transactions on Instrumentation and Measurement}\ }\textbf
  {\bibinfo {volume} {48}},\ \bibinfo {pages} {567} (\bibinfo {year}
  {1999})}\BibitemShut {NoStop}%
\bibitem [{\citenamefont {Galli}\ \emph {et~al.}(2011)\citenamefont {Galli},
  \citenamefont {Bartalini}, \citenamefont {Borri}, \citenamefont {Cancio},
  \citenamefont {Mazzotti}, \citenamefont {{De Natale}},\ and\ \citenamefont
  {Giusfredi}}]{Galli2011}%
  \BibitemOpen
  \bibfield  {author} {\bibinfo {author} {\bibfnamefont {I.}~\bibnamefont
  {Galli}}, \bibinfo {author} {\bibfnamefont {S.}~\bibnamefont {Bartalini}},
  \bibinfo {author} {\bibfnamefont {S.}~\bibnamefont {Borri}}, \bibinfo
  {author} {\bibfnamefont {P.}~\bibnamefont {Cancio}}, \bibinfo {author}
  {\bibfnamefont {D.}~\bibnamefont {Mazzotti}}, \bibinfo {author}
  {\bibfnamefont {P.}~\bibnamefont {{De Natale}}}, \ and\ \bibinfo {author}
  {\bibfnamefont {G.}~\bibnamefont {Giusfredi}},\ }\href {\doibase
  10.1103/PhysRevLett.107.270802} {\bibfield  {journal} {\bibinfo  {journal}
  {Phys. Rev. Lett.}\ }\textbf {\bibinfo {volume} {107}},\ \bibinfo {pages}
  {270802} (\bibinfo {year} {2011})}\BibitemShut {NoStop}%
\bibitem [{\citenamefont {Bressel}, \citenamefont {Ernsting},\ and\
  \citenamefont {Schiller}(2012)}]{Bressel2012}%
  \BibitemOpen
  \bibfield  {author} {\bibinfo {author} {\bibfnamefont {U.}~\bibnamefont
  {Bressel}}, \bibinfo {author} {\bibfnamefont {I.}~\bibnamefont {Ernsting}}, \
  and\ \bibinfo {author} {\bibfnamefont {S.}~\bibnamefont {Schiller}},\ }\href
  {http://www.ncbi.nlm.nih.gov/pubmed/22378438} {\bibfield  {journal} {\bibinfo
   {journal} {Optics letters}\ }\textbf {\bibinfo {volume} {37}},\ \bibinfo
  {pages} {918} (\bibinfo {year} {2012})}\BibitemShut {NoStop}%
\bibitem [{\citenamefont {Ricciardi}\ \emph {et~al.}(2012)\citenamefont
  {Ricciardi}, \citenamefont {{De Tommasi}}, \citenamefont {Maddaloni},
  \citenamefont {Mosca}, \citenamefont {Rocco}, \citenamefont {Zondy},
  \citenamefont {{De Rosa}},\ and\ \citenamefont {{De
  Natale}}}]{Ricciardi2012}%
  \BibitemOpen
  \bibfield  {author} {\bibinfo {author} {\bibfnamefont {I.}~\bibnamefont
  {Ricciardi}}, \bibinfo {author} {\bibfnamefont {E.}~\bibnamefont {{De
  Tommasi}}}, \bibinfo {author} {\bibfnamefont {P.}~\bibnamefont {Maddaloni}},
  \bibinfo {author} {\bibfnamefont {S.}~\bibnamefont {Mosca}}, \bibinfo
  {author} {\bibfnamefont {A.}~\bibnamefont {Rocco}}, \bibinfo {author}
  {\bibfnamefont {J.-J.}\ \bibnamefont {Zondy}}, \bibinfo {author}
  {\bibfnamefont {M.}~\bibnamefont {{De Rosa}}}, \ and\ \bibinfo {author}
  {\bibfnamefont {P.}~\bibnamefont {{De Natale}}},\ }\href {\doibase
  10.1364/OE.20.009178} {\bibfield  {journal} {\bibinfo  {journal} {Opt.
  Express}\ }\textbf {\bibinfo {volume} {20}},\ \bibinfo {pages} {9178}
  (\bibinfo {year} {2012})}\BibitemShut {NoStop}%
\bibitem [{\citenamefont {Schliesser}, \citenamefont {Picqu\'{e}},\ and\
  \citenamefont {H\"{a}nsch}(2012)}]{Schliesser2012}%
  \BibitemOpen
  \bibfield  {author} {\bibinfo {author} {\bibfnamefont {A.}~\bibnamefont
  {Schliesser}}, \bibinfo {author} {\bibfnamefont {N.}~\bibnamefont
  {Picqu\'{e}}}, \ and\ \bibinfo {author} {\bibfnamefont {T.~W.}\ \bibnamefont
  {H\"{a}nsch}},\ }\href {\doibase 10.1038/nphoton.2012.142} {\bibfield
  {journal} {\bibinfo  {journal} {Nat. Phot.}\ }\textbf {\bibinfo {volume}
  {6}},\ \bibinfo {pages} {440} (\bibinfo {year} {2012})}\BibitemShut {NoStop}%
\bibitem [{\citenamefont {Myers}\ \emph {et~al.}(2002)\citenamefont {Myers},
  \citenamefont {Williams}, \citenamefont {Taubman}, \citenamefont {Gmachl},
  \citenamefont {Capasso}, \citenamefont {Sivco}, \citenamefont {Baillargeon},\
  and\ \citenamefont {Cho}}]{Myers2002}%
  \BibitemOpen
  \bibfield  {author} {\bibinfo {author} {\bibfnamefont {T.~L.}\ \bibnamefont
  {Myers}}, \bibinfo {author} {\bibfnamefont {R.~M.}\ \bibnamefont {Williams}},
  \bibinfo {author} {\bibfnamefont {M.~S.}\ \bibnamefont {Taubman}}, \bibinfo
  {author} {\bibfnamefont {C.}~\bibnamefont {Gmachl}}, \bibinfo {author}
  {\bibfnamefont {F.}~\bibnamefont {Capasso}}, \bibinfo {author} {\bibfnamefont
  {D.~L.}\ \bibnamefont {Sivco}}, \bibinfo {author} {\bibfnamefont {J.~N.}\
  \bibnamefont {Baillargeon}}, \ and\ \bibinfo {author} {\bibfnamefont {a.~Y.}\
  \bibnamefont {Cho}},\ }\href {http://www.ncbi.nlm.nih.gov/pubmed/18007745}
  {\bibfield  {journal} {\bibinfo  {journal} {Optics letters}\ }\textbf
  {\bibinfo {volume} {27}},\ \bibinfo {pages} {170} (\bibinfo {year}
  {2002})}\BibitemShut {NoStop}%
\bibitem [{\citenamefont {Bartalini}\ \emph {et~al.}(2010)\citenamefont
  {Bartalini}, \citenamefont {Borri}, \citenamefont {Cancio}, \citenamefont
  {Castrillo}, \citenamefont {Galli}, \citenamefont {Giusfredi}, \citenamefont
  {Mazzotti}, \citenamefont {Gianfrani},\ and\ \citenamefont {{De
  Natale}}}]{Bartalini2010}%
  \BibitemOpen
  \bibfield  {author} {\bibinfo {author} {\bibfnamefont {S.}~\bibnamefont
  {Bartalini}}, \bibinfo {author} {\bibfnamefont {S.}~\bibnamefont {Borri}},
  \bibinfo {author} {\bibfnamefont {P.}~\bibnamefont {Cancio}}, \bibinfo
  {author} {\bibfnamefont {a.}~\bibnamefont {Castrillo}}, \bibinfo {author}
  {\bibfnamefont {I.}~\bibnamefont {Galli}}, \bibinfo {author} {\bibfnamefont
  {G.}~\bibnamefont {Giusfredi}}, \bibinfo {author} {\bibfnamefont
  {D.}~\bibnamefont {Mazzotti}}, \bibinfo {author} {\bibfnamefont
  {L.}~\bibnamefont {Gianfrani}}, \ and\ \bibinfo {author} {\bibfnamefont
  {P.}~\bibnamefont {{De Natale}}},\ }\href {\doibase
  10.1103/PhysRevLett.104.083904} {\bibfield  {journal} {\bibinfo  {journal}
  {Physical Review Letters}\ }\textbf {\bibinfo {volume} {104}},\ \bibinfo
  {pages} {083904} (\bibinfo {year} {2010})}\BibitemShut {NoStop}%
\bibitem [{\citenamefont {Tombez}\ \emph {et~al.}(2011)\citenamefont {Tombez},
  \citenamefont {Francesco}, \citenamefont {Schilt}, \citenamefont {Domenico},
  \citenamefont {Faist}, \citenamefont {Thomann}, \citenamefont {Hofstetter},
  \citenamefont {{Di Francesco}}, \citenamefont {Schilt}, \citenamefont {{Di
  Domenico}}, \citenamefont {Faist}, \citenamefont {Thomann},\ and\
  \citenamefont {Hofstetter}}]{Tombez2011}%
  \BibitemOpen
  \bibfield  {author} {\bibinfo {author} {\bibfnamefont {L.}~\bibnamefont
  {Tombez}}, \bibinfo {author} {\bibfnamefont {J.~D.}\ \bibnamefont
  {Francesco}}, \bibinfo {author} {\bibfnamefont {S.}~\bibnamefont {Schilt}},
  \bibinfo {author} {\bibfnamefont {G.~D.}\ \bibnamefont {Domenico}}, \bibinfo
  {author} {\bibfnamefont {J.}~\bibnamefont {Faist}}, \bibinfo {author}
  {\bibfnamefont {P.}~\bibnamefont {Thomann}}, \bibinfo {author} {\bibfnamefont
  {D.}~\bibnamefont {Hofstetter}}, \bibinfo {author} {\bibfnamefont
  {J.}~\bibnamefont {{Di Francesco}}}, \bibinfo {author} {\bibfnamefont
  {S.}~\bibnamefont {Schilt}}, \bibinfo {author} {\bibfnamefont
  {G.}~\bibnamefont {{Di Domenico}}}, \bibinfo {author} {\bibfnamefont
  {J.}~\bibnamefont {Faist}}, \bibinfo {author} {\bibfnamefont
  {P.}~\bibnamefont {Thomann}}, \ and\ \bibinfo {author} {\bibfnamefont
  {D.}~\bibnamefont {Hofstetter}},\ }\href {<Go to ISI>://000293890800032}
  {\bibfield  {journal} {\bibinfo  {journal} {Opt. Lett.}\ }\textbf {\bibinfo
  {volume} {36}},\ \bibinfo {pages} {3109} (\bibinfo {year}
  {2011})}\BibitemShut {NoStop}%
\bibitem [{\citenamefont {Bartalini}\ \emph {et~al.}(2011)\citenamefont
  {Bartalini}, \citenamefont {Borri}, \citenamefont {Galli}, \citenamefont
  {Giusfredi}, \citenamefont {Mazzotti}, \citenamefont {Edamura}, \citenamefont
  {Akikusa}, \citenamefont {Yamanishi},\ and\ \citenamefont {{De
  Natale}}}]{Bartalini2011}%
  \BibitemOpen
  \bibfield  {author} {\bibinfo {author} {\bibfnamefont {S.}~\bibnamefont
  {Bartalini}}, \bibinfo {author} {\bibfnamefont {S.}~\bibnamefont {Borri}},
  \bibinfo {author} {\bibfnamefont {I.}~\bibnamefont {Galli}}, \bibinfo
  {author} {\bibfnamefont {G.}~\bibnamefont {Giusfredi}}, \bibinfo {author}
  {\bibfnamefont {D.}~\bibnamefont {Mazzotti}}, \bibinfo {author}
  {\bibfnamefont {T.}~\bibnamefont {Edamura}}, \bibinfo {author} {\bibfnamefont
  {N.}~\bibnamefont {Akikusa}}, \bibinfo {author} {\bibfnamefont
  {M.}~\bibnamefont {Yamanishi}}, \ and\ \bibinfo {author} {\bibfnamefont
  {P.}~\bibnamefont {{De Natale}}},\ }\href {<Go to ISI>://000294781200029}
  {\bibfield  {journal} {\bibinfo  {journal} {Opt. Express}\ }\textbf {\bibinfo
  {volume} {19}},\ \bibinfo {pages} {17996} (\bibinfo {year}
  {2011})}\BibitemShut {NoStop}%
\bibitem [{\citenamefont {Mills}\ \emph {et~al.}(2012)\citenamefont {Mills},
  \citenamefont {Gatti}, \citenamefont {Jiang}, \citenamefont {Mohr},
  \citenamefont {Mefford}, \citenamefont {Gianfrani}, \citenamefont {Fermann},
  \citenamefont {Hartl},\ and\ \citenamefont {Marangoni}}]{Mills2012}%
  \BibitemOpen
  \bibfield  {author} {\bibinfo {author} {\bibfnamefont {A.~A.}\ \bibnamefont
  {Mills}}, \bibinfo {author} {\bibfnamefont {D.}~\bibnamefont {Gatti}},
  \bibinfo {author} {\bibfnamefont {J.}~\bibnamefont {Jiang}}, \bibinfo
  {author} {\bibfnamefont {C.}~\bibnamefont {Mohr}}, \bibinfo {author}
  {\bibfnamefont {W.}~\bibnamefont {Mefford}}, \bibinfo {author} {\bibfnamefont
  {L.}~\bibnamefont {Gianfrani}}, \bibinfo {author} {\bibfnamefont
  {M.}~\bibnamefont {Fermann}}, \bibinfo {author} {\bibfnamefont
  {I.}~\bibnamefont {Hartl}}, \ and\ \bibinfo {author} {\bibfnamefont
  {M.}~\bibnamefont {Marangoni}},\ }\href {<Go to ISI>://000309542900053}
  {\bibfield  {journal} {\bibinfo  {journal} {Opt. Lett.}\ }\textbf {\bibinfo
  {volume} {37}},\ \bibinfo {pages} {4083} (\bibinfo {year}
  {2012})}\BibitemShut {NoStop}%
\bibitem [{\citenamefont {Bielsa}\ \emph {et~al.}(2007)\citenamefont {Bielsa},
  \citenamefont {Douillet}, \citenamefont {Valenzuela}, \citenamefont {Karr},\
  and\ \citenamefont {Hilico}}]{Bielsa2007}%
  \BibitemOpen
  \bibfield  {author} {\bibinfo {author} {\bibfnamefont {F.}~\bibnamefont
  {Bielsa}}, \bibinfo {author} {\bibfnamefont {A.}~\bibnamefont {Douillet}},
  \bibinfo {author} {\bibfnamefont {T.}~\bibnamefont {Valenzuela}}, \bibinfo
  {author} {\bibfnamefont {J.-P.}\ \bibnamefont {Karr}}, \ and\ \bibinfo
  {author} {\bibfnamefont {L.}~\bibnamefont {Hilico}},\ }\href {<Go to
  ISI>://000248144600015} {\bibfield  {journal} {\bibinfo  {journal} {Opt.
  Lett.}\ }\textbf {\bibinfo {volume} {32}},\ \bibinfo {pages} {1641} (\bibinfo
  {year} {2007})}\BibitemShut {NoStop}%
\bibitem [{\citenamefont {Bielsa}\ \emph {et~al.}(2008)\citenamefont {Bielsa},
  \citenamefont {Djerroud}, \citenamefont {Goncharov}, \citenamefont
  {Douillet}, \citenamefont {Valenzuela}, \citenamefont {Daussy}, \citenamefont
  {Hilico},\ and\ \citenamefont {Amy-Klein}}]{Bielsa2008}%
  \BibitemOpen
  \bibfield  {author} {\bibinfo {author} {\bibfnamefont {F.}~\bibnamefont
  {Bielsa}}, \bibinfo {author} {\bibfnamefont {K.}~\bibnamefont {Djerroud}},
  \bibinfo {author} {\bibfnamefont {A.}~\bibnamefont {Goncharov}}, \bibinfo
  {author} {\bibfnamefont {A.}~\bibnamefont {Douillet}}, \bibinfo {author}
  {\bibfnamefont {T.}~\bibnamefont {Valenzuela}}, \bibinfo {author}
  {\bibfnamefont {C.}~\bibnamefont {Daussy}}, \bibinfo {author} {\bibfnamefont
  {L.}~\bibnamefont {Hilico}}, \ and\ \bibinfo {author} {\bibfnamefont
  {A.}~\bibnamefont {Amy-Klein}},\ }\href {\doibase 10.1016/j.jms.2007.10.003}
  {\bibfield  {journal} {\bibinfo  {journal} {J. Mol. Spec.}\ }\textbf
  {\bibinfo {volume} {247}},\ \bibinfo {pages} {41} (\bibinfo {year}
  {2008})}\BibitemShut {NoStop}%
\bibitem [{\citenamefont {Cappelli}\ \emph {et~al.}(2012)\citenamefont
  {Cappelli}, \citenamefont {Galli}, \citenamefont {Borri}, \citenamefont
  {Giusfredi}, \citenamefont {Cancio}, \citenamefont {Mazzotti}, \citenamefont
  {Montori}, \citenamefont {Akikusa}, \citenamefont {Yamanishi}, \citenamefont
  {Bartalini},\ and\ \citenamefont {{De Natale}}}]{Cappelli2012}%
  \BibitemOpen
  \bibfield  {author} {\bibinfo {author} {\bibfnamefont {F.}~\bibnamefont
  {Cappelli}}, \bibinfo {author} {\bibfnamefont {I.}~\bibnamefont {Galli}},
  \bibinfo {author} {\bibfnamefont {S.}~\bibnamefont {Borri}}, \bibinfo
  {author} {\bibfnamefont {G.}~\bibnamefont {Giusfredi}}, \bibinfo {author}
  {\bibfnamefont {P.}~\bibnamefont {Cancio}}, \bibinfo {author} {\bibfnamefont
  {D.}~\bibnamefont {Mazzotti}}, \bibinfo {author} {\bibfnamefont
  {A.}~\bibnamefont {Montori}}, \bibinfo {author} {\bibfnamefont
  {N.}~\bibnamefont {Akikusa}}, \bibinfo {author} {\bibfnamefont
  {M.}~\bibnamefont {Yamanishi}}, \bibinfo {author} {\bibfnamefont
  {S.}~\bibnamefont {Bartalini}}, \ and\ \bibinfo {author} {\bibfnamefont
  {P.}~\bibnamefont {{De Natale}}},\ }\href {<Go to ISI>://000311943900002}
  {\bibfield  {journal} {\bibinfo  {journal} {Opt. Lett.}\ }\textbf {\bibinfo
  {volume} {37}},\ \bibinfo {pages} {4811} (\bibinfo {year}
  {2012})}\BibitemShut {NoStop}%
\bibitem [{\citenamefont {Borri}\ \emph {et~al.}(2012)\citenamefont {Borri},
  \citenamefont {Galli}, \citenamefont {Cappelli}, \citenamefont {Bismuto},
  \citenamefont {Bartalini}, \citenamefont {Cancio}, \citenamefont {Giusfredi},
  \citenamefont {Mazzotti}, \citenamefont {Faist},\ and\ \citenamefont {{De
  Natale}}}]{Borri2012}%
  \BibitemOpen
  \bibfield  {author} {\bibinfo {author} {\bibfnamefont {S.}~\bibnamefont
  {Borri}}, \bibinfo {author} {\bibfnamefont {I.}~\bibnamefont {Galli}},
  \bibinfo {author} {\bibfnamefont {F.}~\bibnamefont {Cappelli}}, \bibinfo
  {author} {\bibfnamefont {A.}~\bibnamefont {Bismuto}}, \bibinfo {author}
  {\bibfnamefont {S.}~\bibnamefont {Bartalini}}, \bibinfo {author}
  {\bibfnamefont {P.}~\bibnamefont {Cancio}}, \bibinfo {author} {\bibfnamefont
  {G.}~\bibnamefont {Giusfredi}}, \bibinfo {author} {\bibfnamefont
  {D.}~\bibnamefont {Mazzotti}}, \bibinfo {author} {\bibfnamefont
  {J.}~\bibnamefont {Faist}}, \ and\ \bibinfo {author} {\bibfnamefont
  {P.}~\bibnamefont {{De Natale}}},\ }\href {<Go to ISI>://000302212200008}
  {\bibfield  {journal} {\bibinfo  {journal} {Opt. Lett.}\ }\textbf {\bibinfo
  {volume} {37}},\ \bibinfo {pages} {1011} (\bibinfo {year}
  {2012})}\BibitemShut {NoStop}%
\bibitem [{\citenamefont {Hansen}\ \emph {et~al.}(2013)\citenamefont {Hansen},
  \citenamefont {Ernsting}, \citenamefont {Vasilyev}, \citenamefont {Grisard},
  \citenamefont {Lallier}, \citenamefont {G\'{e}rard},\ and\ \citenamefont
  {Schiller}}]{Hansen2013}%
  \BibitemOpen
  \bibfield  {author} {\bibinfo {author} {\bibfnamefont {M.~G.}\ \bibnamefont
  {Hansen}}, \bibinfo {author} {\bibfnamefont {I.}~\bibnamefont {Ernsting}},
  \bibinfo {author} {\bibfnamefont {S.~V.}\ \bibnamefont {Vasilyev}}, \bibinfo
  {author} {\bibfnamefont {A.}~\bibnamefont {Grisard}}, \bibinfo {author}
  {\bibfnamefont {E.}~\bibnamefont {Lallier}}, \bibinfo {author} {\bibfnamefont
  {B.}~\bibnamefont {G\'{e}rard}}, \ and\ \bibinfo {author} {\bibfnamefont
  {S.}~\bibnamefont {Schiller}},\ }\href
  {http://www.ncbi.nlm.nih.gov/pubmed/24216928} {\bibfield  {journal} {\bibinfo
   {journal} {Optics express}\ }\textbf {\bibinfo {volume} {21}},\ \bibinfo
  {pages} {27043} (\bibinfo {year} {2013})}\BibitemShut {NoStop}%
\bibitem [{\citenamefont {Galli}\ \emph {et~al.}(2013)\citenamefont {Galli},
  \citenamefont {{Siciliani de Cumis}}, \citenamefont {Cappelli}, \citenamefont
  {Bartalini}, \citenamefont {Mazzotti}, \citenamefont {Borri}, \citenamefont
  {Montori}, \citenamefont {Akikusa}, \citenamefont {Yamanishi}, \citenamefont
  {Giusfredi}, \citenamefont {Cancio},\ and\ \citenamefont {{De
  Natale}}}]{Galli2013}%
  \BibitemOpen
  \bibfield  {author} {\bibinfo {author} {\bibfnamefont {I.}~\bibnamefont
  {Galli}}, \bibinfo {author} {\bibfnamefont {M.}~\bibnamefont {{Siciliani de
  Cumis}}}, \bibinfo {author} {\bibfnamefont {F.}~\bibnamefont {Cappelli}},
  \bibinfo {author} {\bibfnamefont {S.}~\bibnamefont {Bartalini}}, \bibinfo
  {author} {\bibfnamefont {D.}~\bibnamefont {Mazzotti}}, \bibinfo {author}
  {\bibfnamefont {S.}~\bibnamefont {Borri}}, \bibinfo {author} {\bibfnamefont
  {A.}~\bibnamefont {Montori}}, \bibinfo {author} {\bibfnamefont
  {N.}~\bibnamefont {Akikusa}}, \bibinfo {author} {\bibfnamefont
  {M.}~\bibnamefont {Yamanishi}}, \bibinfo {author} {\bibfnamefont
  {G.}~\bibnamefont {Giusfredi}}, \bibinfo {author} {\bibfnamefont
  {P.}~\bibnamefont {Cancio}}, \ and\ \bibinfo {author} {\bibfnamefont
  {P.}~\bibnamefont {{De Natale}}},\ }\href {\doibase 10.1063/1.4799284}
  {\bibfield  {journal} {\bibinfo  {journal} {Appl. Phys. Lett.}\ }\textbf
  {\bibinfo {volume} {102}},\ \bibinfo {pages} {121117} (\bibinfo {year}
  {2013})}\BibitemShut {NoStop}%
\bibitem [{\citenamefont {Stoeffler}\ \emph {et~al.}(2011)\citenamefont
  {Stoeffler}, \citenamefont {Darqui\'{e}}, \citenamefont {Shelkovnikov},
  \citenamefont {Daussy}, \citenamefont {Amy-Klein}, \citenamefont
  {Chardonnet}, \citenamefont {Guy}, \citenamefont {Crassous}, \citenamefont
  {Huet}, \citenamefont {Soulard},\ and\ \citenamefont
  {Asselin}}]{Stoeffler2011}%
  \BibitemOpen
  \bibfield  {author} {\bibinfo {author} {\bibfnamefont {C.}~\bibnamefont
  {Stoeffler}}, \bibinfo {author} {\bibfnamefont {B.}~\bibnamefont
  {Darqui\'{e}}}, \bibinfo {author} {\bibfnamefont {A.}~\bibnamefont
  {Shelkovnikov}}, \bibinfo {author} {\bibfnamefont {C.}~\bibnamefont
  {Daussy}}, \bibinfo {author} {\bibfnamefont {A.}~\bibnamefont {Amy-Klein}},
  \bibinfo {author} {\bibfnamefont {C.}~\bibnamefont {Chardonnet}}, \bibinfo
  {author} {\bibfnamefont {L.}~\bibnamefont {Guy}}, \bibinfo {author}
  {\bibfnamefont {J.}~\bibnamefont {Crassous}}, \bibinfo {author}
  {\bibfnamefont {T.~R.}\ \bibnamefont {Huet}}, \bibinfo {author}
  {\bibfnamefont {P.}~\bibnamefont {Soulard}}, \ and\ \bibinfo {author}
  {\bibfnamefont {P.}~\bibnamefont {Asselin}},\ }\href {\doibase
  10.1039/c0cp01806f} {\bibfield  {journal} {\bibinfo  {journal} {Phys. Chem.
  Chem. Phys.}\ }\textbf {\bibinfo {volume} {13}},\ \bibinfo {pages} {854}
  (\bibinfo {year} {2011})}\BibitemShut {NoStop}%
\bibitem [{\citenamefont {Lemarchand}\ \emph {et~al.}(2010)\citenamefont
  {Lemarchand}, \citenamefont {Djerroud}, \citenamefont {Darqui\'{e}},
  \citenamefont {Lopez}, \citenamefont {Amy-Klein}, \citenamefont {Chardonnet},
  \citenamefont {Bord\'{e}}, \citenamefont {Briaudeau},\ and\ \citenamefont
  {Daussy}}]{Lemarchand2010}%
  \BibitemOpen
  \bibfield  {author} {\bibinfo {author} {\bibfnamefont {C.}~\bibnamefont
  {Lemarchand}}, \bibinfo {author} {\bibfnamefont {K.}~\bibnamefont
  {Djerroud}}, \bibinfo {author} {\bibfnamefont {B.}~\bibnamefont
  {Darqui\'{e}}}, \bibinfo {author} {\bibfnamefont {O.}~\bibnamefont {Lopez}},
  \bibinfo {author} {\bibfnamefont {A.}~\bibnamefont {Amy-Klein}}, \bibinfo
  {author} {\bibfnamefont {C.}~\bibnamefont {Chardonnet}}, \bibinfo {author}
  {\bibfnamefont {C.~J.}\ \bibnamefont {Bord\'{e}}}, \bibinfo {author}
  {\bibfnamefont {S.}~\bibnamefont {Briaudeau}}, \ and\ \bibinfo {author}
  {\bibfnamefont {C.}~\bibnamefont {Daussy}},\ }\href {\doibase
  10.1007/s10765-010-0755-3} {\bibfield  {journal} {\bibinfo  {journal}
  {International Journal of Thermophysics}\ }\textbf {\bibinfo {volume} {31}},\
  \bibinfo {pages} {1347} (\bibinfo {year} {2010})}\BibitemShut {NoStop}%
\bibitem [{\citenamefont {Bernard}\ \emph {et~al.}(1995)\citenamefont
  {Bernard}, \citenamefont {Durand}, \citenamefont {George}, \citenamefont
  {Nicolaisen}, \citenamefont {Amy-Klein},\ and\ \citenamefont
  {Chardonnet}}]{Bernard1995}%
  \BibitemOpen
  \bibfield  {author} {\bibinfo {author} {\bibfnamefont {V.}~\bibnamefont
  {Bernard}}, \bibinfo {author} {\bibfnamefont {P.~E.}\ \bibnamefont {Durand}},
  \bibinfo {author} {\bibfnamefont {T.}~\bibnamefont {George}}, \bibinfo
  {author} {\bibfnamefont {H.~W.}\ \bibnamefont {Nicolaisen}}, \bibinfo
  {author} {\bibfnamefont {A.}~\bibnamefont {Amy-Klein}}, \ and\ \bibinfo
  {author} {\bibfnamefont {C.}~\bibnamefont {Chardonnet}},\ }\href@noop {}
  {\bibfield  {journal} {\bibinfo  {journal} {IEEE Journal of Quantum
  Electronics}\ }\textbf {\bibinfo {volume} {QE-31}},\ \bibinfo {pages} {1913}
  (\bibinfo {year} {1995})}\BibitemShut {NoStop}%
\bibitem [{\citenamefont {Elliott}, \citenamefont {Roy},\ and\ \citenamefont
  {Smith}(1982)}]{Elliott1982}%
  \BibitemOpen
  \bibfield  {author} {\bibinfo {author} {\bibfnamefont {D.}~\bibnamefont
  {Elliott}}, \bibinfo {author} {\bibfnamefont {R.}~\bibnamefont {Roy}}, \ and\
  \bibinfo {author} {\bibfnamefont {S.}~\bibnamefont {Smith}},\ }\href
  {\doibase 10.1103/PhysRevA.26.12} {\bibfield  {journal} {\bibinfo  {journal}
  {Phys. Rev. A}\ }\textbf {\bibinfo {volume} {26}},\ \bibinfo {pages} {12}
  (\bibinfo {year} {1982})}\BibitemShut {NoStop}%
\bibitem [{\citenamefont {Bishof}\ \emph {et~al.}(2013)\citenamefont {Bishof},
  \citenamefont {Zhang}, \citenamefont {Martin},\ and\ \citenamefont
  {Ye}}]{Bishof2013}%
  \BibitemOpen
  \bibfield  {author} {\bibinfo {author} {\bibfnamefont {M.}~\bibnamefont
  {Bishof}}, \bibinfo {author} {\bibfnamefont {X.}~\bibnamefont {Zhang}},
  \bibinfo {author} {\bibfnamefont {M.~J.}\ \bibnamefont {Martin}}, \ and\
  \bibinfo {author} {\bibfnamefont {J.}~\bibnamefont {Ye}},\ }\href {\doibase
  10.1103/PhysRevLett.111.093604} {\bibfield  {journal} {\bibinfo  {journal}
  {Physical Review Letters}\ }\textbf {\bibinfo {volume} {111}},\ \bibinfo
  {pages} {093604} (\bibinfo {year} {2013})}\BibitemShut {NoStop}%
\bibitem [{\citenamefont {Zhu}\ and\ \citenamefont {Hall}(1993)}]{Zhu1993}%
  \BibitemOpen
  \bibfield  {author} {\bibinfo {author} {\bibfnamefont {M.}~\bibnamefont
  {Zhu}}\ and\ \bibinfo {author} {\bibfnamefont {J.~L.}\ \bibnamefont {Hall}},\
  }\href {\doibase 10.1364/JOSAB.10.000802} {\bibfield  {journal} {\bibinfo
  {journal} {J. Opt. Soc. Am. B}\ }\textbf {\bibinfo {volume} {10}},\ \bibinfo
  {pages} {802} (\bibinfo {year} {1993})}\BibitemShut {NoStop}%
\bibitem [{\citenamefont {Chardonnet}(1989)}]{Chardonnet1989}%
  \BibitemOpen
  \bibfield  {author} {\bibinfo {author} {\bibfnamefont {C.}~\bibnamefont
  {Chardonnet}},\ }\emph {\bibinfo {title} {{Spectroscopie de saturation de
  hautes pr\'{e}cision et sensibilit\'{e} en champ laser fort. Applications aux
  mol\'{e}cules OsO$_4$, SF$_6$ et CO$_2$ et \`{a} la m\'{e}trologie des
  fr\'{e}quences}}},\ \href@noop {} {Ph.D. thesis},\ \bibinfo  {school}
  {Universit\'{e} Paris 13}, \bibinfo {address} {Villetaneuse} (\bibinfo {year}
  {1989})\BibitemShut {NoStop}%
\bibitem [{\citenamefont {Gu\'{e}na}\ \emph {et~al.}(2012)\citenamefont
  {Gu\'{e}na}, \citenamefont {Abgrall}, \citenamefont {Rovera}, \citenamefont
  {Laurent}, \citenamefont {Chupin}, \citenamefont {Lours}, \citenamefont
  {Santarelli}, \citenamefont {Rosenbusch}, \citenamefont {Tobar},
  \citenamefont {Li}, \citenamefont {Gibble}, \citenamefont {Clairon},\ and\
  \citenamefont {Bize}}]{Guena2012}%
  \BibitemOpen
  \bibfield  {author} {\bibinfo {author} {\bibfnamefont {J.}~\bibnamefont
  {Gu\'{e}na}}, \bibinfo {author} {\bibfnamefont {M.}~\bibnamefont {Abgrall}},
  \bibinfo {author} {\bibfnamefont {D.}~\bibnamefont {Rovera}}, \bibinfo
  {author} {\bibfnamefont {P.}~\bibnamefont {Laurent}}, \bibinfo {author}
  {\bibfnamefont {B.}~\bibnamefont {Chupin}}, \bibinfo {author} {\bibfnamefont
  {M.}~\bibnamefont {Lours}}, \bibinfo {author} {\bibfnamefont
  {G.}~\bibnamefont {Santarelli}}, \bibinfo {author} {\bibfnamefont
  {P.}~\bibnamefont {Rosenbusch}}, \bibinfo {author} {\bibfnamefont
  {M.}~\bibnamefont {Tobar}}, \bibinfo {author} {\bibfnamefont
  {R.}~\bibnamefont {Li}}, \bibinfo {author} {\bibfnamefont {K.}~\bibnamefont
  {Gibble}}, \bibinfo {author} {\bibfnamefont {A.}~\bibnamefont {Clairon}}, \
  and\ \bibinfo {author} {\bibfnamefont {S.}~\bibnamefont {Bize}},\ }\href
  {\doibase 10.1109/TUFFC.2012.2208} {\bibfield  {journal} {\bibinfo  {journal}
  {IEEE transactions on ultrasonics, ferroelectrics, and frequency control}\
  }\textbf {\bibinfo {volume} {59}},\ \bibinfo {pages} {391} (\bibinfo {year}
  {2012})}\BibitemShut {NoStop}%
\bibitem [{\citenamefont {Chanteau}\ \emph {et~al.}(2013)\citenamefont
  {Chanteau}, \citenamefont {Lopez}, \citenamefont {Zhang}, \citenamefont
  {Nicolodi}, \citenamefont {Argence}, \citenamefont {Auguste}, \citenamefont
  {Abgrall}, \citenamefont {Chardonnet}, \citenamefont {Santarelli},
  \citenamefont {Darqui\'{e}}, \citenamefont {{Le Coq}},\ and\ \citenamefont
  {Amy-Klein}}]{Chanteau2013}%
  \BibitemOpen
  \bibfield  {author} {\bibinfo {author} {\bibfnamefont {B.}~\bibnamefont
  {Chanteau}}, \bibinfo {author} {\bibfnamefont {O.}~\bibnamefont {Lopez}},
  \bibinfo {author} {\bibfnamefont {W.}~\bibnamefont {Zhang}}, \bibinfo
  {author} {\bibfnamefont {D.}~\bibnamefont {Nicolodi}}, \bibinfo {author}
  {\bibfnamefont {B.}~\bibnamefont {Argence}}, \bibinfo {author} {\bibfnamefont
  {F.}~\bibnamefont {Auguste}}, \bibinfo {author} {\bibfnamefont
  {M.}~\bibnamefont {Abgrall}}, \bibinfo {author} {\bibfnamefont
  {C.}~\bibnamefont {Chardonnet}}, \bibinfo {author} {\bibfnamefont
  {G.}~\bibnamefont {Santarelli}}, \bibinfo {author} {\bibfnamefont
  {B.}~\bibnamefont {Darqui\'{e}}}, \bibinfo {author} {\bibfnamefont
  {Y.}~\bibnamefont {{Le Coq}}}, \ and\ \bibinfo {author} {\bibfnamefont
  {A.}~\bibnamefont {Amy-Klein}},\ }\href {\doibase
  10.1088/1367-2630/15/7/073003} {\bibfield  {journal} {\bibinfo  {journal}
  {New J. Phys.}\ }\textbf {\bibinfo {volume} {15}},\ \bibinfo {pages} {073003}
  (\bibinfo {year} {2013})}\BibitemShut {NoStop}%
\end{thebibliography}%

\end{document}